\documentclass[aps,pra,amsfonts,showpacs]{revtex4}

\usepackage{pstricks}

\renewcommand{\>}{\rangle}

\newcommand{\be}{\begin{equation}}
\newcommand{\ee}{\end{equation}}
\newcommand{\bea}{\begin{eqnarray}}
\newcommand{\eea}{\end{eqnarray}}

\begin{document}

\title{Operator Quantum Error Correcting Subsystems for Self-Correcting Quantum Memories}
\author{Dave Bacon}
\affiliation{Department of Computer Science \& Engineering, University of
Washington, Seattle, WA 98195}
\date{June 6, 2005}

\begin{abstract}
The most general method for encoding quantum information is not to encode the
information into a subspace of a Hilbert space, but to encode information into
a subsystem of a Hilbert space.  Recently this notion has led to a more general
notion of quantum error correction known as operator quantum error correction.
In standard quantum error correcting codes, one requires the ability to apply a
procedure which exactly reverses on the error correcting subspace any
correctable error. In contrast, for operator error correcting subsystems, the
correction procedure need not undo the error which has occurred, but instead
one must perform correction only modulo the subsystem structure.  This does not
lead to codes which differ from subspace codes, but does lead to recovery
routines which explicitly make use of the subsystem structure.  Here we present
two examples of such operator error correcting subsystems.  These examples are
motivated by simple spatially local Hamiltonians on square and cubic lattices.
In three dimensions we provide evidence, in the form a simple mean field
theory, that our Hamiltonian gives rise to a system which is self-correcting.
Such a system will be a natural high-temperature quantum memory, robust to
noise without external intervening quantum error correction procedures.
\end{abstract}

\pacs{03.67.Lx,03.67.–a,75.10.Pq}
 \maketitle

\section{Introduction}
In the early days of quantum computation, the analog nature of quantum
information and quantum transforms, as well as the effect of noise processes on
quantum systems, were thought to pose severe
obstacles\cite{Unruh:95a,Landauer:96a} towards the experimental realization of
the exponential speedups promised by quantum computers over classical
computers\cite{Bernstein:93a,Simon:94a,Shor:94a,Raz:99a}.  Soon, however, a
remarkable theory of fault-tolerant quantum
computation\cite{Shor:95a,Steane:96b,Knill:97a,Bennett:96a,Knill:98a,Knill:98b,Shor:96a,Aharonov:97a,Gottesman:98a,Kitaev:03a,Preskill:98a}
emerged which dealt with these problems and showed that quantum computers are
indeed more similar to probabilistic classical computers than to analog
devices.  Analog computers have a computational power which is dependent on a
lack of noise and on exponential precision, whereas probabilistic classical
computers can be error corrected and made effectively digital even in the
presence of noise and non-exponential precision.  The theory of fault-tolerant
quantum computation establishes that quantum computers are truly digital
devices deserving of the moniker computer. An essential idea in the development
of the theory of fault-tolerant quantum computation was the notion that quantum
information could be encoded into
subspaces\cite{Shor:95a,Steane:96b,Knill:97a,Bennett:96a} (quantum error
correcting codes) and thereafter protected from degradation via active
procedures of detection and correction of errors. Encoding quantum information
into subspaces, however, is not the most general method of encoding quantum
information into a quantum system.  The most general notion for encoding
quantum information is to encode the information into a subsystem of the
quantum system\cite{Knill:97a,Knill:00a,Viola:01a}. This has been perhaps best
exploited in the theory of noiseless
subsystems\cite{Knill:00a,Kempe:01a,Bacon:01a,Zanardi:01a,Zanardi:03a,Shabani:05a}
and dynamic recoupling schemes\cite{Viola:00a,Zanardi:01a,Wu:02a}.  Recently a
very general notion of quantum error correction has appeared under the moniker
of ``operator quantum error correction.''\cite{Kribs:05a,Kribs:05b}  In this
work the possibility of encoding into subsystems for active error correction is
explicitly examined. While it was found\cite{Kribs:05a,Kribs:05b} that the
notion of a encoding into a subsystem does not lead to new codes (all subsystem
codes could be thought of as arising from subspace codes), encoding into a
subsystem {\em does} lead to different recovery procedures for quantum
information which has been encoded into a subsystem. Hence operator quantum
error correcting codes, while not offering the hope of more general codes, do
offer the possibility of new quantum error correcting routines, and in
particular to the possibility of codes which might help improve the threshold
for fault-tolerant quantum computation due to the lessened complexity of the
error correcting routine.

In this paper we present two examples of operator quantum error correcting
codes which use subsystem encodings.  The codes we present have the interesting
property that the recovery routine does not restore information encoded into a
subspace, but recovers the information encoded into a subsystem.  Using the
$[n,d,k]$ labelling a quantum error correcting codes, where $n$ is the number
of qubits used in the code, $d$ is the distance of the code, and $k$ is the
number of encoded qubits for the code, our codes are $[n^2,n,1]$ and
$[n^3,n,1]$ quantum error correcting codes.  The subsystem structure of our
codes is explicitly exploited in the recovery routine for the code, and because
of this they are substantially simpler than any subspace code derived from
these codes.

While the two codes we present are interesting in there own right, there is a
further motivation for these codes above and beyond their exploitation of the
subsystem structure in the recovery routine.  The two operator quantum error
correcting subsystems we present are motivated by two interesting Hamiltonians
defined on two and three-dimensional square and cubic lattices of qubits with
certain anisotropic spin-spin interactions\cite{Bacon:01a}.  The
three-dimensional version of this system is particularly intriguing since it
offers the possibility of being a self-correcting quantum memory.  In a
self-correcting quantum memory, quantum error correction is enacted not by the
external control of a complicated quantum error correction scheme, but instead
by the natural physics of the device. Such a quantum memory offers the
possibility for removing the need for a quantum microarchitecture to perform
quantum error correction and could therefore profoundly speed up the process of
building a quantum computer. In this paper we present evidence, in the form of
a simple mean field argument, that the three-dimensional system we consider is
a self-correcting quantum memory. We also show that the operator error
correcting subsystem structure of this code is an important component to not
only the self-correcting properties of this system, but also to encoding and
decoding information in this system.

The organization of the paper is as follows.  In Section~\ref{sec:subsystem} we
review the notion of encoding information into a subsystem and discuss the
various ways in which this has been applied to noiseless subsystems and dynamic
recoupling methods for protecting quantum information.  Next, in
Section~\ref{sec:qecs}, we discuss how operator error correcting subsystems
work and how they differ from standard quantum error correcting codes.  Our
first example of a operator quantum error correcting subsystem is presented in
Section~\ref{sec:2d} where we introduce an example on a square lattice.  The
second, and more interesting, example of a operator quantum error correcting
subsystem is given in Section~\ref{sec:3d} where we discuss an example on a
cubic lattice.  In Section~\ref{sec:hamiltonian} we introduction the notion of
a self-correcting quantum memory and present arguments that a particular
Hamiltonian on a cubic lattice related to our cubic lattice subsystem is
self-correcting.  We conclude in Section~\ref{sec:conc} with a discussion of
open problems and the prospects for operator quantum error correcting
subsystems and self-correcting quantum memories.

\section{Subsystem Encoding} \label{sec:subsystem}

Consider two qubits.  The Hilbert space of these qubits is given by
$\mathbb{C}^2 \otimes \mathbb{C}^2$.  Pick some fiducial basis for each qubit
labelled by $|0\>$ and $|1\>$.  One way to encode a single qubit of information
into these two qubits is to encode the information into a subspace of the joint
system.  For example, we can define the logical basis states $|0_L\>={1 \over
\sqrt{2}}(|01\>-|10\>)$ and $|1_L\>=|11\>$ such that a single qubit of
information can be encoded as $\alpha|0_L\> + \beta |1_L\>= {\alpha \over
\sqrt{2}}(|01\>-|10\>)+\beta|11\>$.  This is an example of idea of encoding
quantum information into a subspace, in this case the subspace spanned by
$|0_L\>$ and $|1_L\>$.  But another way to encode a single qubit of information
is to encode this information into one of the two qubits.  In particular if we
prepare the state $|\psi\> \otimes (\alpha|0\> + \beta|1\>)$ for an arbitrary
single qubit state $|\psi\>$, then we have also encoded a single qubit of
information in our system.  This time, however, we have encoded in the
information into a subsystem of the system.  It is important to note that the
subsystem encoding works for an arbitrary state $|\psi\rangle$.  If we fix
$|\psi\rangle$ to some known state, then we are again encoding into a subspace.
We reserve the nomenclature of ``encoding into a subsystem'' to times in which
$|\psi\rangle$ is arbitrary.

More generally, if we have some Hilbert space ${\mathcal H}$, then a subsystem
${\mathcal C}$ is a Hilbert space arising from ${\mathcal H}$
as\cite{Knill:97a,Knill:00a}
\begin{equation}
{\mathcal H}=({\mathcal C} \otimes {\mathcal D}) \oplus {\mathcal E}.
\end{equation}
Here we have taken our Hilbert space and partitioned it into two subspaces,
${\mathcal E}$ and a subspace perpendicular to ${\mathcal E}$.  On this
perpendicular subspaces, we have introduced a tensor product structure,
${\mathcal C} \otimes {\mathcal D}$.  We can then encode information into the
first subsystem ${\mathcal C}$.  This can be achieved by preparing the quantum
information we wish to encode $\rho_C$ into the first subsystem, ${\mathcal C}$
along with any arbitrary state $\rho_D$ into the second subsystem ${\mathcal
D}$:
\be
\rho=(\rho_C \otimes \rho_D) \oplus 0.
\ee

The fact that quantum information can most generally be encoded into a
subsystem was an essential insight used in the construction of noiseless
(decoherence-free)
subsystems\cite{Knill:00a,Kempe:01a,Bacon:01a,Zanardi:01a,Zanardi:03a,Shabani:05a}.
Suppose we have a system with Hilbert space ${\mathcal H}_S$ and an environment
with Hilbert space ${\mathcal H}_E$.  The coupling between these two systems
will be described by an interaction Hamiltonian $H_{int}$ which acts on the
tensor product of these two spaces ${\mathcal H}_S \otimes {\mathcal H}_B$. The
idea of a noiseless subsystem is that it is often the case that there is a
symmetry of the system-environment interaction such that the action of the
interaction Hamiltonian factors with respect to some subsystem structure on the
system's Hilbert space,
\begin{equation}
H_{int}=\sum_\alpha \left [\left( I_{d} \otimes D_\alpha \right) \oplus
E_\alpha \right] \otimes B_\alpha,
\end{equation}
where $I_d$ is the d-dimensional identity operator acting on the subsystem code
space ${\mathcal C}$, $D_\alpha$ acts on the subsystem ${\mathcal D}$,
$E_\alpha$ acts on the orthogonal subspace ${\mathcal E}$, and $B_\alpha$
operates on the environment Hilbert space ${\mathcal H}_E$. When our
interaction Hamiltonian possesses a symmetry leading to such a structure, then,
if we encode quantum information into ${\mathcal C}$, this information will not
be effected by the system-environment coupling.  Thus information encoded in
such a subsystem will be protected from the effect of decoherence and hence
exists in a noiseless subsystem.  Noiseless subsystems were a generalization of
decoherence-free subspaces\cite{Zanardi:97a,Lidar:98a}, this latter idea
occurring when the subsystem structure is not exploited, ${\mathcal D}={\mathbb
C}$, and then encoding quantum information is simply encoding quantum
information into a subspace.  Subsystems have also been used in dynamic
recoupling techniques\cite{Viola:00a,Zanardi:01a,Wu:02a} where symmetries are
produced by an active symmetrization of the system's component of the
system-environment evolution.

\section{Operator Quantum Error Correcting Subsystems} \label{sec:qecs}

Here we examine the implications of encoding information into a subsystem for
quantum error correcting protocols\cite{Kribs:05a,Kribs:05b}.  Suppose that we
encode quantum information into a subsystem ${\mathcal C}$ of some quantum
system with full Hilbert space ${\mathcal H}=({\mathcal C} \otimes {\mathcal
D}) \oplus {\mathcal E}$.  Now suppose some quantum operation (corresponding to
an error) occurs on our system. Following the standard quantum error correcting
paradigm, we then apply a recovery procedure to the system.  When ${\mathcal
D}={\mathbb C}$, i.e. when we are encoding into a quantum error correcting
subspace, then the quantum error correcting criteria is simply that the effect
of the error process followed by the recovery operation should act as identity
on this subspace. If we encode information into a subsystem, however, this
criteria is changed to {\em only} requiring that the recovery operation should
act as identity on the subsystem ${\mathcal C}$.  In particular we do not care
if the effect of an error followed by our recovery procedure enacts some
nontrivial procedure on the ${\mathcal D}$ subsystem.  In fact our error
correcting procedure may induce some nontrivial action on the ${\mathcal D}$
subsystem in the process of restoring information encoded in the ${\mathcal C}$
subsystem.

How does the above observation modify the basic theory of quantum error
correcting codes?  In standard quantum error correction, we encode into some
error correcting subspace with basis $|i\rangle$.  The necessary and sufficient
condition for there to be a procedure under which quantum information can be
restored under a given set of errors $E_a$ is given by
\be
\langle i | E_a^\dagger E_b |j\rangle =\delta_{i,j} c_{a,b}. \label{eq:qecs}
\ee
For the case of encoding into a subsystem this necessary and sufficient
condition is modified as follows. Let $|i\rangle \otimes |k\rangle$ denote a
basis for the subspace ${\mathcal C} \otimes {\mathcal D}$.  Then Kribs {\em et
al.}\cite{Kribs:05a,Kribs:05b} showed a necessary
condition\cite{Knill:97a,Bennett:96a} for the quantum error correcting is given
by
\be
\left(\langle i | \otimes \langle k| \right)E_a^\dagger E_b\left( | j \rangle
\otimes |l\rangle \right) =\delta_{i,j} m_{a,b,k,l}. \label{eq:qecsub}
\ee
That this condition is also sufficient has recently also been shown by Nielsen
and Poulin\cite{Nielsen:05a}.  As noted in \cite{Kribs:05a,Kribs:05b}, a code
constructed from the subsystem operator quantum error correcting criteria can
always be used to construct a subspace code which satisfies the subspace
criteria Eq.~(\ref{eq:qecs}).  We note, however, that while this implies that
the notion of using subsystems for quantum error correction does not lead to
new quantum error correcting codes above and beyond subspace encodings, the
codes constructed which exploit the subsystem structure have error recovery
routines which are distinct from those which arise when encoding into a
subspace.  In particular, when one encodes into a subsystem, the recovery
routine does not need to fix errors which occur on other subsystems.  Below we
will present examples of subsystem encodings in which the subsystem structure
of the encoding is essential not for the existence of the quantum error
correcting properties, but it essential for the simple recovery routine we
present.

\section{Two-Dimensional Operator Quantum Error Correcting Subsystem} \label{sec:2d}

Here we construct an operator quantum error correcting subsystem for a code
which lives on a two-dimensional square lattice.  This code makes explicit use
of the subsystem structure in its error recovery procedure.  A familiarity with
the stabilizer formalism for quantum error correcting codes is assumed (see
\cite{Gottesman:97a,Nielsen:00a} for overviews.)

\subsection{Preliminary Definitions}

Consider a square lattice of size $n \times n$ with qubits located at the
vertices of this lattice.  Let $O_{i,j}$ denote the operator $O$ acting on the
qubit located at the $i$th row and $j$th column of this lattice tensored with
identity on all other qubits.  Recall that the Pauli operators on a single
qubit are
\begin{eqnarray}
X=\left[\begin{array}{cc} 0 & 1 \\ 1 & 0 \end{array} \right],
iY=\left[\begin{array}{cc} 0 & -1 \\ 1 & 0 \end{array} \right],~{\rm and}~
Z=\left[\begin{array}{cc} 1 & 0 \\ 0 & -1 \end{array} \right].
\end{eqnarray}
It is convenient to use a compact notation to denote Pauli operator on our
$n^2$ qubits by using two $n^2$ bit strings,
\begin{equation}
P(a,b)=\prod_{i,j=1}^n X_{i,j}^{a_{i,j}} Z_{i,j}^{b_{i,j}}= \prod_{i,j=1}^n \left\{%
\begin{array}{ll}
    X_{i,j} & {\rm if}~a_{i,j}=1~{\rm and}~b_{i,j}=0 \\
    Z_{i,j} & {\rm if}~a_{i,j}=0~{\rm and}~b_{i,j}=1 \\
    -iY_{i,j} &{\rm if}~a_{i,j}=1~{\rm and}~b_{i,j}=1
\end{array}%
\right.,
\end{equation}
where $a,b \in {\mathbb Z}_2^{n^2}$ are $n$ by $n$ matrices of bits.  Together
with a phase, $i^\phi$, $\phi \in {\mathbb Z}_4$, a generic element of the
Pauli group on our $n^2$ qubits is given by $i^\phi P(a,b)$.  We will often
refer to a Pauli operator as being made up of $X$ and $Z$ operators, noting
that when both appear, the actual Pauli operator is the $iY$ operator.

We begin by defining three sets of operators which are essential to
understanding the subsystem structure of our qubits.  Each of these sets will
be made up of Pauli operators.

\begin{figure}[h]
\begin{pspicture}(0,0)(6,6)
  \qline(1,1)(5,1)
  \qline(1,1.5)(5,1.5)
  \qline(1,2)(5,2)
  \qline(1,2.5)(5,2.5)
  \qline(1,3)(5,3)
  \qline(1,3.5)(5,3.5)
  \qline(1,4)(5,4)
  \qline(1,4.5)(5,4.5)
  \qline(1,5)(5,5)
  \qline(1,1)(1,5)
  \qline(1.5,1)(1.5,5)
  \qline(2,1)(2,5)
  \qline(2.5,1)(2.5,5)
  \qline(3,1)(3,5)
  \qline(3.5,1)(3.5,5)
  \qline(4,1)(4,5)
  \qline(4.5,1)(4.5,5)
  \qline(5,1)(5,5)
  \rput[bl](2.9,2.9){$X$}
  \rput[bl](2.9,3.4){$X$}
  \rput[bl](0.9,2.4){$Z$}
  \rput[bl](1.4,2.4){$Z$}
  \rput[bl](3.9,3.9){$Y$}
  \rput[bl](4.4,3.9){$Y$}
  \rput[bl](4.4,4.4){$X$}
  \rput[bl](3.9,3.4){$X$}
  \rput[bl](1.4,1.4){$Z$}
  \rput[bl](3.4,1.4){$Z$}
 \psframe[linestyle=dashed](1.25,1.25)(3.75,1.75)
 \psframe[linestyle=dashed](0.75,2.25)(1.75,2.75)
 \psframe[linestyle=dashed](2.75,2.75)(3.25,3.75)
 \psframe[linestyle=dashed](3.75,3.25)(4.75,4.75)
\end{pspicture}
\caption{Above we have represented elements of the ${\mathcal T}$.  Each set of
operators enclosed in a rectangle represents a Pauli operator acting on the
particular qubits tensored with the identity on all other qubits.  Each of the
operators enclosed in the dotted rectangles are elements of ${\mathcal T}$.}
\label{fig:t}
\end{figure}
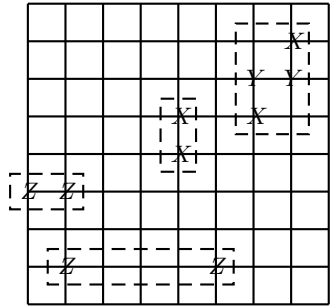

The first set of Pauli operators which will concern us, ${\mathcal T}$, is made
up of Pauli operators which have an even number of $X_{i,j}$ operators in each
column and an even number of $Z_{i,j}$ operators in each row:
\bea
{\mathcal T}= \left \{ (-1)^\phi P(a,b) | \phi \in {\mathbb
Z}_2,\bigoplus_{i=1}^n a_{i,j}=0,~{\rm and} \bigoplus_{j=1}^n b_{i,j}=0 \right
\},
\eea
where $\bigoplus$ denotes the binary exclusive-or operation (we use it
interchangeably with the direct sum operation with context distinguishing the
two uses.) Note that these operators form a group under multiplication. This
group can be generated by nearest neighbor operators on our cubic lattice
\be
{\mathcal T} = \left < X_{i,j}X_{i+1,j},Z_{j,i}Z_{j,i+1}, \forall i \in
{\mathbb Z}_{n-1},j \in {\mathbb Z}_{n}\right>.
\ee
Examples of elements of the group ${\mathcal T}$ are diagrammed in
Fig.~\ref{fig:t}.

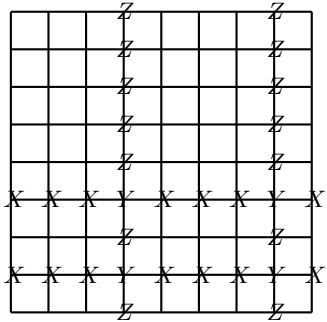
\begin{figure}[h]
\begin{pspicture}(0,0)(6,6)
  \qline(1,1)(5,1)
  \qline(1,1.5)(5,1.5)
  \qline(1,2)(5,2)
  \qline(1,2.5)(5,2.5)
  \qline(1,3)(5,3)
  \qline(1,3.5)(5,3.5)
  \qline(1,4)(5,4)
  \qline(1,4.5)(5,4.5)
  \qline(1,5)(5,5)
  \qline(1,1)(1,5)
  \qline(1.5,1)(1.5,5)
  \qline(2,1)(2,5)
  \qline(2.5,1)(2.5,5)
  \qline(3,1)(3,5)
  \qline(3.5,1)(3.5,5)
  \qline(4,1)(4,5)
  \qline(4.5,1)(4.5,5)
  \qline(5,1)(5,5)
  \rput[bl](0.9,2.4){$X$}
  \rput[bl](1.4,2.4){$X$}
  \rput[bl](1.9,2.4){$X$}
  \rput[bl](2.9,2.4){$X$}
  \rput[bl](3.4,2.4){$X$}
  \rput[bl](3.9,2.4){$X$}
  \rput[bl](4.9,2.4){$X$}
  \rput[bl](0.9,1.4){$X$}
  \rput[bl](1.4,1.4){$X$}
  \rput[bl](1.9,1.4){$X$}
  \rput[bl](2.9,1.4){$X$}
  \rput[bl](3.4,1.4){$X$}
  \rput[bl](3.9,1.4){$X$}
  \rput[bl](4.9,1.4){$X$}
  \rput[bl](2.4,0.9){$Z$}
  \rput[bl](2.4,1.4){$Y$}
  \rput[bl](2.4,1.9){$Z$}
  \rput[bl](2.4,2.4){$Y$}
  \rput[bl](2.4,2.9){$Z$}
  \rput[bl](2.4,3.4){$Z$}
  \rput[bl](2.4,3.9){$Z$}
  \rput[bl](2.4,4.4){$Z$}
  \rput[bl](2.4,4.9){$Z$}
  \rput[bl](4.4,0.9){$Z$}
  \rput[bl](4.4,1.4){$Y$}
  \rput[bl](4.4,1.9){$Z$}
  \rput[bl](4.4,2.4){$Y$}
  \rput[bl](4.4,2.9){$Z$}
  \rput[bl](4.4,3.4){$Z$}
  \rput[bl](4.4,3.9){$Z$}
  \rput[bl](4.4,4.4){$Z$}
  \rput[bl](4.4,4.9){$Z$}

\end{pspicture}
\caption{A nontrivial element of the group ${\mathcal S}$.  This element has an
even number of columns which are entirely $X$ operators multiplied by an even
number of rows which are entirely $Y$ operators.  Notice how the $Y$ elements
appear where both of these conditions are met.}\label{fig:s}
\end{figure}

The second set of Pauli operators we will be interested in is a subset of
${\mathcal T}$, which we denote ${\mathcal S}$.  ${\mathcal S}$ consists of
Pauli operators which are made up of an even number of rows consisting entirely
of $X$ operators and an even number of columns consisting entirely of $Z$
operators,
\be
{\mathcal S}=\left\{ P(a,b)| \bigoplus_{i=1}^n \left(\bigwedge_{j=1}^n
a_{i,j}\right)=0, \bigoplus_{j=1}^n \left(\bigwedge_{i=1}^n b_{i,j}\right)=0
\right\}
\ee
where $\bigwedge$ is the binary and operation.  ${\mathcal S}$ is also a group.
In fact it is an Abelian subgroup of ${\mathcal T}$.  Further all of the
elements of ${\mathcal S}$ commute not just with each other, but with all of
the elements of ${\mathcal T}$.  It can be generated by nearest row and column
operators,
\begin{equation}
{\mathcal S}= \left< \prod_{i=1}^n X_{j,i} X_{j+1,i}, \prod_{i=1}^n Z_{i,j}
Z_{i,j+1}, \forall j \in {\mathbb Z}_{n-1}\right>.
\end{equation}
These generators will be particularly important for us, so we will denote them
by
\bea
S_i^X=\prod_{j=1}^n X_{i,j} X_{i+1,j},~{\rm and}~S_j^Z=\prod_{i=1}^n Z_{i,j}
Z_{i,j+1}.
\eea
${\mathcal S}$ is a stabilizer group familiar from the standard theory of
quantum error correcting codes.  An example of an element in ${\mathcal S}$ is
given in Fig.~\ref{fig:s}

\begin{figure}[h]
\begin{pspicture}(0,0)(6,6)
  \qline(1,1)(5,1)
  \qline(1,1.5)(5,1.5)
  \qline(1,2)(5,2)
  \qline(1,2.5)(5,2.5)
  \qline(1,3)(5,3)
  \qline(1,3.5)(5,3.5)
  \qline(1,4)(5,4)
  \qline(1,4.5)(5,4.5)
  \qline(1,5)(5,5)
  \qline(1,1)(1,5)
  \qline(1.5,1)(1.5,5)
  \qline(2,1)(2,5)
  \qline(2.5,1)(2.5,5)
  \qline(3,1)(3,5)
  \qline(3.5,1)(3.5,5)
  \qline(4,1)(4,5)
  \qline(4.5,1)(4.5,5)
  \qline(5,1)(5,5)
  \rput[bl](0.9,2.4){$X$}
  \rput[bl](1.4,2.4){$X$}
  \rput[bl](1.9,2.4){$X$}
  \rput[bl](2.9,2.4){$X$}
  \rput[bl](3.4,2.4){$X$}
  \rput[bl](3.9,2.4){$X$}
  \rput[bl](4.4,2.4){$X$}
  \rput[bl](4.9,2.4){$X$}
  \rput[bl](2.4,0.9){$Z$}
  \rput[bl](2.4,1.4){$Z$}
  \rput[bl](2.4,1.9){$Z$}
  \rput[bl](2.4,2.4){$Y$}
  \rput[bl](2.4,2.9){$Z$}
  \rput[bl](2.4,3.4){$Z$}
  \rput[bl](2.4,3.9){$Z$}
  \rput[bl](2.4,4.4){$Z$}
  \rput[bl](2.4,4.9){$Z$}
\end{pspicture}
\caption{A nontrivial element of the set ${\mathcal L}$.  This element has an
odd number of columns which are entirely $X$ operators multiplied by an
oddnumber of rows which are entirely $Y$ operators.  It represents an encoded
$Y$ operator on the encoded qubit as described in
Sec.~\ref{sec:log}}\label{fig:l}
\end{figure}
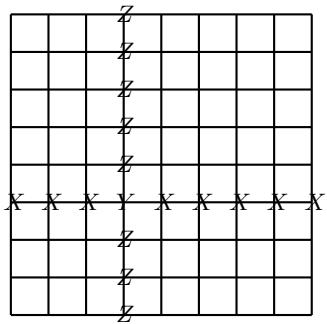

The final set of operators which we will consider, ${\mathcal L}$, is similar
to ${\mathcal S}$ except that the evenness condition becomes an oddness
condition,
\bea
{\mathcal L}=\left\{ (-1)^\phi P(a,b)|\phi \in {\mathbb Z}_2, \bigoplus_{i=1}^n
\left(\bigwedge_{j=1}^n a_{i,j}\right)=1,\bigoplus_{j=1}^n
\left(\bigwedge_{i=1}^n b_{i,j}\right)=1\right\}.
\eea
This set does not by itself form a group, but together with ${\mathcal S}$ it
does form a group. This combined group is not Abelian.  ${\mathcal L}$ has the
property that all of its elements commute with those of ${\mathcal T}$ and
${\mathcal S}$.  A nontrivial element of ${\mathcal L}$ is given in
Fig.~\ref{fig:l}.

\subsection{Subsystem Structure}

We will now elucidate how ${\mathcal T}$, ${\mathcal S}$, and ${\mathcal L}$
are related to a subsystem structure on our $n^2$ qubits.  Let ${\mathcal
H}=({\mathbb C}^2)^{n^2}$ denote the Hilbert space of our $n^2$ qubits.

We first note that since ${\mathcal S}$ consists of a set of mutually commuting
observables, we can use these observables to label subspaces of ${\mathcal H}$.
In particular we can label these subspaces by the $2(n-1)$ $\pm 1$-valued
eigenvalues of the $S_i^X$ and $S_i^Z$ operators.  Let us denote these
eigenvalues by $s_i^X$ and $s_i^Z$ respectively and the length $n-1$ string of
these $\pm 1$ eigenvalues by $s^X$ and $s^Z$. We can thus decompose ${\mathcal
H}$ into subspaces as
\be
{\mathcal H}=\bigoplus_{s_1^X,\dots,s_{n-1}^X,s_1^Z,\dots,s_{n-1}^Z = \pm 1}
{\mathcal H}_{s^X,s^Z}=\bigoplus_{s^X,s^Z} {\mathcal H}_{s^X,s^Z}.
\ee
By standard arguments in the stabilizer formalism, each of the ${\mathcal
H}_{s^X,s_Z}$ subspaces is of dimension $d=2^{n^2-2(n-1)}$.  Just for
completeness, we note that the operators $S_i^X$ and $S_i^Z$ act under this
decomposition as
\bea
S_i^X&=&\bigoplus_{s^X,s^Z} s_i^X I_{2^{n^2-2(n-1)}}, \nonumber \\
S_i^Z&=&\bigoplus_{s^X,s^Z} s_i^Z I_{2^{n^2-2(n-1)}}.
\eea

Now examine the two groups ${\mathcal T}$ and the group generated by elements
of ${\mathcal L}$ and ${\mathcal S}$. Both of these groups are non-Abelian.
All of the elements of ${\mathcal T}$ and ${\mathcal L}$ commute with elements
of ${\mathcal S}$. Further, all of the elements of ${\mathcal T}$ and
${\mathcal L}$ commute with each other. This implies, via Schur's
lemma\cite{Cornwell:97a,James:01a}, that ${\mathcal L}$ and ${\mathcal T}$ must
be represented on ${\mathcal H}_{s^X,s^Z}$ by a subsystem action.  In
particular the full Hilbert space splits as
\begin{equation}
{\mathcal H}=\bigoplus_{s^X,s^Z} {\mathcal H}^{\mathcal T}_{s^X,s^Z} \otimes
{\mathcal H}^{\mathcal L}_{s^X,s^Z}, \label{eq:sub2d}
\end{equation}
such that operators from $T \in {\mathcal T}$ act on the first tensor product
\be
T=\bigoplus_{s^X,s^Z} T_{s^X,s^Z} \otimes I_2, \forall T \in {\mathcal T},
\ee
and the operators from $L \in {\mathcal L}$ act on the second tensor product
\be
L=\bigoplus_{s^X,s^Z} I_{2^{(n-1)^2}} \otimes L_{s^X,s^Z}, \forall L \in
{\mathcal L}.
\ee
Here we have assigned dimensions $2^{(n-1)^2}$ and $2$ to these tensor product
spaces.  To see why these dimensionalities arise we appeal to the stabilizer
formalism.  We note that modulo the stabilizer structure of ${\mathcal S}$,
${\mathcal L}$ is a single encoded qubit.  Similarly if one examines the
following set of $(n-1)^2$ operators from ${\mathcal T}$,
\bea
\bar{Z}_{i,j} = Z_{i,j}Z_{i,j+1}, \quad \bar{X}_{i,j} = \prod_{k=1}^j
X_{i,k}X_{n-1,k},
\eea
where $i \in {\mathbb Z}_{n-1}, j \in {\mathbb Z}_{n-1}$, one finds that modulo
the stabilizer they are equivalent to $(n-1)^2$ encoded Pauli operators.

The subsystem code we now propose encodes a single qubit into the Hilbert space
${\mathcal H}_{s^X,s^Z}^{\mathcal L}$ with $s^X_i=s^X_i=+1, \forall i \in
{\mathbb Z}_{n-1}$ (choices with other $\pm 1$ choices form an equivalent code
in the same way that stabilizer codes can be chosen for different stabilizer
generator eigenvalues.) The code we propose thus encodes one qubit of quantum
information into a subsystem of the $n^2$ ``bare'' qubits.  We stress that the
encoding we perform is truly a subsystem encoding: we do not care what the
state of the ${\mathcal H}_{s^X,s^Z}^{\mathcal T}$ subsystem is.  For
simplicity it may be possible to begin by encoding into a subspace which
includes our particular subsystem (i.e. by fixing the state on ${\mathcal
H}_{s^X,s^Z}^{\mathcal T}$) but this encoding is not necessary and indeed,
after our recover routine for the information encoded into ${\mathcal
H}_{s^X,s^Z}^{\mathcal L}$, we will not know the state of the ${\mathcal
H}_{s^X,s^Z}^{\mathcal T}$ subsystem.  We will denote the Hilbert space
${\mathcal H}_{s^X,s^Z}^{\mathcal L}$ with $s^X_i=s^X_i=+1, \forall i \in
{\mathbb Z}_{n-1}$ by ${\mathcal H}_{s^X=s^Z=\{+1\}^{n-1}}^{\mathcal L}$.

\subsection{Subsystem Error Correcting Procedure}

If we encode quantum information into the subsystem ${\mathcal
H}_{s^X=s^Z=\{+1\}^{n-1}}^{\mathcal L}$, then what sort of error correcting
properties does this encoding result in?  We will see that the ${\mathcal S}$
operators can be used to perform an error correcting procedure which restores
the information on ${\mathcal H}_{s^X=s^Z=\{+1\}^{n-1}}^{\mathcal L}$, but
which often acts nontrivially on the subsystem ${\mathcal
H}_{s^X=s^Z=\{+1\}^{n-1}}^{\mathcal T}$.  This exploitation of the subsystem
structure in the correction procedure is what distinguishes our subsystem
operator quantum error correcting code from standard subspace quantum error
correcting codes.

Suppose that a Pauli error $P(a,b)$ occurs on our system.  For a Pauli operator
$P(a,b)$, define the following error strings:
\begin{equation}
e_j(a)=\bigoplus_{i=1}^{n} a_{i,j}, \quad f_i(b)=\bigoplus_{j=1}^n b_{i,j}.
\end{equation}
Notice that if $e_j=f_i=0,\forall i,j$, then this implies that $P(a,b)$ is in
the set ${\mathcal T}$.  Further note that in this case, the effect of $P(a,b)$
is to only act on the ${\mathcal H}_{s^X,s^Z}^{\mathcal T}$ subsystems, i.e.
$P(a,b)$ is block diagonal under our subsystem decomposition,
Eq.~(\ref{eq:sub2d}), acting as
\be
P(a,b)=\bigoplus_{s^X,s^Z} E_{s^X,s^Z}(a,b) \otimes I_2.
\ee
where $E_{s^X,s^Z}(a,b)$ is a nontrivial operator depending on the subspace
labels $s^X,s^Z$ and the type of Pauli error $(a,b)$. Therefore errors of this
form ($e_j=f_i=0$) do not cause errors on our information encoded in ${\mathcal
H}_{s^X=s^Z=\{+1\}^{n-1}}^{\mathcal L}$. With respect to the errors of this
form, the information is encoded into a noiseless subsystem\cite{Knill:00a}.

Returning now to the case of a general $P(a,b)$, from the above argument we see
that if we can apply a Pauli operator $Q(c,d)$ such that $Q(c,d)P(a,b)$ is a
new error, call it $R(a^\prime,b^\prime)$, which has error strings
$e_j^\prime(a^\prime)=f_i^\prime(b^\prime)=0,\forall i,j$, then we will have a
procedure for fixing the error $P(a,b)$, {\em modulo the subsystem structure of
our encoded quantum information}.  In other words, our error correcting
procedure need not result in producing the identity action on the subspace
labelled by $s^X_i=s^Z_i=+1, \forall i \in {\mathbb Z}_{n-1}$, but need only
produce the identity action on the subsystem ${\mathcal
H}_{s^X=s^Z=\{+1\}^{n-1}}^{\mathcal L}$.  We can perform just such a procedure
by using the elements of ${\mathcal S}$ as a syndrome for which errors of small
enough size can be corrected.

To see how this works, suppose $P(a,b)$ occurs on our system.  Then note that
measuring $S_i^X$ is equivalent to determining
\be
\bigoplus_{j=1}^n (b_{i,j} \oplus b_{i+1,j})=f_i(b) \oplus f_{i+1}(b),
\ee
and similarly measuring $S_j^Z$ is equivalent to determining
\be
\bigoplus_{i=1}^n (a_{i,j} \oplus a_{i,j+1})=e_j(a) \oplus e_{j+1}(a).
\ee
Note that all $2(n-1)$ of these measurements can be performed simultaneously
since the elements of ${\mathcal S}$ all commute with each other.  We wish to
use these measurement outcomes to restore the system to $e_j(a)=f_i(b)=0$ (if
possible.)

To see how to do this, treat the $f_i(b)$ as a $n$ bit codeword for a simple
redundancy code (i.e. the two codewords are $f_i(b)=0,\forall i$ and $f_i(b)=1,
\forall i$).  A similar procedure will hold for the $e_i(b)$. Measuring the
$n-1$ operators $S_i^X$ is equivalent to measuring the syndrome of our
redundancy code.  In particular we can use this syndrome to apply an error
correcting procedure for the $f_i(b)$ bit strings. The result of this
correction procedure is to restore the system to either the codeword
$f_i(b)=0,\forall i$ or the codeword $f_i(b)=1,\forall i$. The former
corresponds to an error correction procedure which can succeed (given that an
equivalent procedure for the $e_j(a)$ bit strings also succeeds), whereas the
latter procedure is one where the error correction procedure will fail. Notice
that our error correcting procedure, when it succeeds, is only guaranteed to
restore the system to $f_i(b)=0$ and $e_j(b)=0$, and thus the full effect of
the procedure may be to apply some nontrivial operator to the ${\mathcal
H}_{s^X,s^Z}^{\mathcal T}$ subsystems.

Let us be more detailed in describing the error correcting procedure for the
$f_i(b)$ code words.  Let $s_i^X$ be the result of our measurements of the
$S_i^X$ operators.  Give the $s_i^X$ we can construct two possible bit strings
$f_i^\prime$ and $\neg f_i^\prime$ ($\neg$ denotes the negation operation)
consistent with these measurements. Let $H(f^\prime)$ and $H(\neg f^\prime)$
denote the Hamming weight of these bit strings (i.e. $H(f^\prime)$ is the
number of $1$s in the $n$ bits $f_i^\prime$) and define $f^{\prime\prime}$ to
be the bit string $f^\prime$ or $\neg f^\prime$ with the smallest Hamming
weight.  We now apply an operation consisting only of $Z_{i,j}$ operators.  In
particular we apply the operator
\be
Q_1(f^{\prime \prime})=\prod_{i=1}^{n} Z_{i,j_0}^{\delta_{f_i^{\prime
\prime},1}},
\ee
for any fixed column index $j_0$.  The operator $Q_1(f^{\prime \prime}) P(a,b)$
is then seen to be of one of two forms: either this new operator has the $Z$
error string equal to all zeros or all ones.  In the first case we have
successfully restored the system to the all $f_i(b)=0$ codeword, whereas for
the second case, we have failed.  How many $Z$ errors can be corrected in this
fashion?  If $P(a,b)$ consisted of $Z$ errors $b$ with an error string $f_i(b)$
with a Hamming weight of this string $H(f)$ which is less than or equal to
$\lfloor {n-1 \over 2} \rfloor$, then the correction procedure will succeed.
Thus the code we have constructed is a $[n^2,n,1]$ code: it encodes a single
qubit into $n^2$ qubits and has a distance $n$.

Above we have focused on the case of Pauli $Z$ errors.  Clearly an analogous
argument holds for Pauli $X$ errors (with the role of the rows and columns
reversed.) Further, Pauli $Y$ errors are taken care of by the combined action
of these two procedures.  By the standard arguments of digitizing errors in
quantum error correcting codes, we have thus shown how our operator quantum
error correcting subsystem code can correct up to $\lfloor {n-1 \over 2}
\rfloor$ arbitrary single qubit errors.

\subsection{Logical Operators}\label{sec:log}

We comment here on the logical operators (operators which act on the encoded
subsystem) for this code. From our analysis of the subsystem structure, it is
clear that elements of ${\mathcal L}$ act on the subsystem.  Thus, for
instance, the effect of a row of Pauli $X$ operators is to enact an encoded
Pauli $X$ operation on the coded subsystem.  We can choose a labelling of the
subsystem such that
\begin{equation}
\bar{X}=\prod_{j=1}^n X_{1,j} = \bigoplus_{s^X,s^Z} {I}_{2^{(n-1)^2}} \otimes
X,
\end{equation}
while in the same basis the effect of a column of Pauli $Z$ operators is to
enact and an encoded Pauli $Z$ on the coded subsystem,
\begin{equation}
\bar{Z} =\prod_{i=1}^n Z_{i,1} = \bigoplus_{s^X,s^Z} {I}_{2^{(n-1)^2}} \otimes
Z.
\end{equation}
These two operators can then be used to enact any Pauli operator on the encoded
quantum information.  Notice that other elements of ${\mathcal L}$ also act as
encoded Pauli operators on ${\mathcal H}_{s^X=s^Z=\{+1\}^{n-1}}^{\mathcal L}$
(but act with differing signs on the other $s^X,s^Z$ labelled subspaces.) An
important property of these logical operators is that they can be enacted by
performing single qubit operators and no coupling between the different qubits
is needed. This is important because it will allow us to assume an independent
error model for error which occur when we imprecisely implement these gates on
our encoded quantum information.  Not only can the above construction be used
to implement the Pauli operators on our subsystem code, it can also be used to
measure the Pauli operators on our subsystem code.

Another easily implementable operation on our code is a logical controlled-not.
Suppose we take two identically sized two-dimensional codes and stack them on
top of each other. Then the application of a transverse controlled-not operator
between all $n^2$ of these two systems will enact a logical controlled-not
between the two encoded qubits. To see this note that if we treat the elements
of the set ${\mathcal S}$ as a stabilizer code, then these transverse operators
preserve the combined stabilizer ${\mathcal S} \times {\mathcal S}$ and that
the action of the $n^2$ controlled-not gates do not mix the ${\mathcal L}
\times {\mathcal L}$ and ${\mathcal T} \times {\mathcal T}$ operators.

Gottesman\cite{Gottesman:98a,Gottesman:97a} has shown that given the ability to
measure and apply the encoded Pauli operators along with the ability to perform
a controlled NOT on a stabilizer code, one can perform any encoded operation
which is in the normalizer of the Pauli group (i.e. the gate set relevant to
the Gottesman-Knill theorem\cite{Gottesman:97a}).

We have seen how to implement encoded Pauli operators and the controlled-NOT on
the information encoded into our subsystem.  These operations do not allow for
universal quantum computation, so an important open question for our subsystem
code is to find an easily implementable method for completing this gateset to a
universal set of gates.

\subsection{Hamiltonian Model of the two-dimensional Subsystem Code}

An interesting offshoot the above two-dimensional operator quantum error
correcting subsystem is the analysis of a particularly simple Hamiltonian whose
ground state has a degeneracy which corresponds to the subsystem code.  We
introduce this Hamiltonian here in order to make our analysis of a similar
Hamiltonian for our three-dimensional subsystem code more transparent. The
Hamiltonian is given by nearest neighbor interactions constructed entirely from
operators in the set ${\mathcal T}$,
\be
H=-\lambda \sum_{i=1}^{n}\sum_{j=1}^{n-1}\left( Z_{i,j}Z_{i,j+1} + X_{j,i}
X_{j+1,i}\right). \label{eq:ham2d}
\ee
Since $H$ is constructed entirely from elements of ${\mathcal T}$, this
Hamiltonian can be decomposed as
\be
H=\bigoplus_{s^X,s^Z} H_{s^X,s^Z} \otimes I_2.
\ee
To understand the exact nature of this Hamiltonian, we would need to
diagonalize each of the $H_{s^X,s^Z}$.  What can be said, however, is that the
ground state of the system will arise as the ground state of one or more of the
$H_{s^X,s^Z}$ (numerical diagonalization of systems with a few qubits show that
the ground state comes from only the $s^X_i=s^Z_i=+1$ subsystem and we
conjecture that this subspace always contains the ground state.) If $k$ of the
$H_{s^X,s^Z}$ contribute to the ground state, the degeneracy of the ground
state will be $2k$ due to the subsystem corresponding to the ${\mathcal L}$
operators.  Thus we see that we can encode quantum information into the
subsystem degeneracy of this Hamiltonian, in the similar manner that
information is encoded into the ground state of a Hamiltonian related to the
toric code\cite{Kitaev:03a,Dennis:02a,Ogburn:99a}.  However, we do not know
whether this system exhibits a gap in its excitation spectrum similar to that
which exists in the toric codes.  In Section~\ref{sec:hamiltonian} we return
will introduce a similar Hamiltonian for our three-dimensional operator quantum
error correcting subsystem.

\section{Three-dimensional Operator Quantum Error Correcting Subsystem} \label{sec:3d}

We now turn to a three-dimensional operator quantum error correcting subsystem
which is a generalization of the two-dimensional subsystem code we presented
above. In particular, whereas the construction for the two-dimensional model
relied on the structure of ${\mathcal T}$ containing Pauli operators with even
number of Pauli $Z$'s in a row and even number of Pauli $X$'s in a column, in
the three-dimensional case we rely on a new set of operators with an even
number of Pauli $Z$'s in the $yz$ plane and an even number of Pauli $X$'s in
the $xy$ plane.

Consider a cubic lattice of size $n \times n \times n$ with qubits located at
the vertices of this lattice and let $n$ be odd.  Let $O_{i,j,k}$ denote the
operator $O$ acting on the qubit located at the $(i,j,k)$th lattice site
tensored with identity on all other qubits. We again use a compact notation to
denote Pauli operator on our $n^3$ qubits by using two $n^3$ bit strings,
\begin{equation}
P(a,b)= \prod_{i,j,k=1}^n X_{i,j,k}^{a_{i,j,k}} Z_{i,j,,k}^{b_{i,j,k}}=\prod_{i,j,k=1}^n \left\{%
\begin{array}{ll}
    X_{i,j,k} & {\rm if}~a_{i,j,k}=1~{\rm and}~b_{i,j,k}=0 \\
    Z_{i,j,k} & {\rm if}~a_{i,j,k}=0~{\rm and}~b_{i,j,k}=1 \\
    -iY_{i,j,k} &{\rm if}~a_{i,j,k}=1~{\rm and}~b_{i,j,k}=1
\end{array}%
\right.,
\end{equation}
where $a,b \in {\mathbb Z}_2^{n^3}$ are $n$ by $n$ by $n$ arrays of bits.

As in the two-dimensional case, we will define three sets of operators,
${\mathcal T}_3$, ${\mathcal S}_3$, and ${\mathcal L}_3$ which are essential to
understanding the subsystem structure of our qubits.

The first set of Pauli operators which will concern us, ${\mathcal T}_3$, is
made up of Pauli operators which have an even number of $X_{i,j,k}$ operators
in each xy-plane and an even number of $Z_{i,j,k}$ operators in each yz-plane:
\be
{\mathcal T}_3= \left\{ (-1)^\phi P(a,b) | \phi \in {\mathbb Z}_2,
\bigoplus_{i,j=1}^n a_{i,j,k}=0,{\rm and}~\bigoplus_{j,k=1}^n b_{i,j,k}=0
\right\}.
\ee
These operators, like the analogous two-dimensional ${\mathcal T}$ form a group
under multiplication. This group can be generated by nearest neighbor operators
on our cubic lattice
\be
{\mathcal T}_3= \left < X_{k,i,j}X_{k+1,i,j},X_{i,k,j}X_{i,k+1,j},
Z_{i,j,k}Z_{i,j,k+1}, Z_{i,k,j}Z_{i,k+1,j}, \forall i,j \in {\mathbb Z}_n,k \in
{\mathbb Z}_{n-1}\right>
\ee

The second set of Pauli operators we will be interested in is a subset of
${\mathcal T}_3$, which we denote ${\mathcal S}_3$.  ${\mathcal S}_3$ consists
of Pauli operators which are made up of an even number of xy-planes made
entirely of Pauli $Z$ operators and an even number of yz-planes made entirely
of Pauli $X$ operators:
\be
{\mathcal S}_3=\left\{ P(a,b)| \bigoplus_{i=1}^n \left(\bigwedge_{j,k=1}^n
a_{i,j,k}\right)=0, \bigoplus_{k=1}^n \left(\bigwedge_{i,j=1}^n
b_{i,j,k}\right)=0\right\}.
\ee
${\mathcal S}_3$ is an Abelian subgroup of ${\mathcal T}_3$ and all of the
elements of ${\mathcal S}_3$ commute with all of the elements of ${\mathcal
T}_3$.  It can be generated by nearest xy-plane and yz-plane operators:
\begin{equation}
{\mathcal S}_3= \left< \prod_{i,j=1}^n X_{i,j,k} X_{i,i,k+1}, \prod_{i,j=1}^n
Z_{k,i,j} Z_{k+1,i,j}, \forall k \in {\mathbb Z}_{n-1}\right>.
\end{equation}
We label these generators, as before:
\bea
S_k^X=\prod_{i,j=1}^n X_{i,j,k} X_{i,j,k},~{\rm and}~S_i^Z=\prod_{j,k=1}^n
Z_{i,j,k} Z_{i+1,j,k}.\label{eq:s3d}
\eea
${\mathcal S}_3$ is again a stabilizer group.

The final set of operators which we will consider, ${\mathcal L}_3$, is similar
to ${\mathcal S}_3$ except that the evenness condition becomes and oddness
condition
\be
{\mathcal L}_3=\left\{ (-1)^\phi P(a,b)|\phi \in {\mathbb Z}_2,
\bigoplus_{i=1}^n \left(\bigwedge_{j,k=1}^n a_{i,j,k}\right)=1,
\bigoplus_{k=1}^n \left(\bigwedge_{i,j=1}^n b_{i,j,k}\right)=1\right\}.
\ee
${\mathcal L}_3$ together with ${\mathcal S}_3$ forms a group and all of the
elements of ${\mathcal L}_3$ commute with those of ${\mathcal T}_3$.

\subsection{Subsystem Structure}

All three of the sets, ${\mathcal T}_3$, ${\mathcal S}_3$, and ${\mathcal L}_3$
will play a directly analogous role to the sets ${\mathcal T}$, ${\mathcal S}$,
and ${\mathcal L}$ in our two-dimensional model.  In particular if we let
${\mathcal H}$ denote the Hilbert space of our $n^3$ qubits, then we can
partition this space into subspaces labelled by the $2(n-1)$ different $\pm 1$
eigenvalues of the operators $S_k^X$ and $S_i^Z$ of Eq.~\ref{eq:s3d}.  Again we
will label these eigenvalues by $s_k^X$ and $s_i^Z$, with $s^X$ and $s^Z$
labelling these strings.  The Hilbert space of the system then decomposes as
\be
{\mathcal H}=\bigoplus_{s_1^X,\dots,s_{n-1}^X,s_1^Z,\dots,s_{n-1}^Z = \pm 1}
{\mathcal H}_{s^X,s^Z}=\bigoplus_{s^X,s^Z} {\mathcal H}_{s^X,s^Z}.
\ee
Again, the ${\mathcal H}_{s^X,s^Z}$ subspaces have a tensor product structure,
${\mathcal H}_{s^X,s^Z}={\mathcal H}^{{\mathcal T}}_{s^X,s^Z} \otimes {\mathcal
H}^{\mathcal T}_{s^X,s^Z}$, such that elements of ${\mathcal L}_3$ act as
\be
L=\bigoplus_{s^X,s^Z} I_{2^{n^3-2n+1}} \otimes L_{s^X,s^Z}, \forall L \in
{\mathcal L}_3,
\ee
and those of ${\mathcal T}_3$ act as
\be
T=\bigoplus_{s^X,s^Z} T_{s^X,s^Z} \otimes I_2, \forall T \in {\mathcal T}_3.
\ee

\subsection{Subsystem Quantum Error Correcting Procedure}

The subsystem error correcting procedure for the three-dimensional code nearly
directly mimics that of the two-dimensional code.  Here we discuss how the
subsystem error correcting procedure works without going into the details as we
did in the two-dimensional case.  The three-dimensional procedure is nearly
identical to that of the two-dimensional procedure with the sets ${\mathcal
T}$, ${\mathcal S}$, and ${\mathcal L}$ interchanged with the sets ${\mathcal
T}_3$, ${\mathcal S}_3$, and ${\mathcal L}_3$ respectively.

We will again encode our quantum information into the ${\mathcal
H}_{s^X=s^Z=\{+1\}^{n-1}}^{\mathcal L}$ subsystem.  Whereas for the
two-dimensional code we defined error strings for the rows and column
conditions of the set ${\mathcal T}$, now we define error strings for the $xy$
and $yz$ plane conditions of the set ${\mathcal T}_3$.  If the Pauli error
$P(a,b)$ occurs on our system, then we can define the two error strings
\be
e_k(a)=\bigoplus_{i,j=1}^n a_{i,j,k}~{\rm and }~f_i(b)=\bigoplus_{j,k=1}^n
b_{i,j,k}.
\ee
Pauli errors with $e_k(a)=f_i(b)=0$ are errors from ${\mathcal T}_3$ and act
trivially on the information encoded into the ${\mathcal
H}_{s^X_k=s^Z_i=+1}^{\mathcal L}$ subsystem.

The quantum error correction procedure is then directly analogous to the one
for the two-dimensional code.  We measure the $s_k^X$ and $s_i^Z$ operators and
treat these as nearest neighbor parity checks for a redundancy code on the
$f_i(b)$ and $e_k(a)$ respectively.  Then in direct analogy with the
two-dimensional code, we can apply a subsystem error correcting procedure which
restores the system modulo the subsystem structure.  The three-dimensional code
is a $[n^3,n,1]$ code.  We have thus gained nothing in terms of the distance of
the code, but, as we will argue in the next Section, the three-dimensional code
when converted to a Hamiltonian whose ground state is the subsystem code has
intriguing features not found in the two-dimensional code.

\subsection{Logical Operators} \label{sec:log3d}

The logical operators for the three-dimensional code are directly analogous to
those in the two-dimensional code.  As in the two-dimensional code, operators
from ${\mathcal L}_3$ can be used to enact Pauli operators on the information
encoded into the ${\mathcal H}_{s^X_k=s^Z_i=+1}^{\mathcal L}$ subsystem.
Similarly, we can enact a controlled-not between two encoded qubits by
performing $n^3$ controlled-not gates between two identical copies of the code.
This allows us to again perform any operation in the normalizer of the Pauli
group on our encoded qubits.

\section{Self-Correction in the Three-dimensional Example}
\label{sec:hamiltonian}

In the 1930s, when Alan Turing wrote his now classic
papers\cite{Turing:36a,Turing:46a} laying out the foundations of computer
science, there was absolutely no reason to believe that any computing device
such as the one described by Turing could actually be built. One of the
foremost problems, immediately apparent to the engineers of the day, was the
lack of reliable components out of which a computer could be built. Von Neumann
solved this problem, in theory, by showing that robust encoding of the
classical information could be used to overcome errors and faulty components in
a computer\cite{vonNeumann:56a}.  Despite Von Neumann's theoretical ideas, it
took the invention of the transistor and the integrated circuit, to mention
only the broadest innovations, in order to bring forth the technological
movement now known as the computer revolution.  The overarching result of the
technological innovations responsible for the computer revolution was the
development of techniques which exhibited Von Neumann's theoretical ideas in a
{\em natural setting}.  Modern computers naturally correct errors in both the
storage and manipulation of classical information.  The task of robust storage
and manipulation of the data is essentially guaranteed by the physics of these
devices. {\em There are distinct physical reasons why robust storage and
manipulation of classical information is possible.}

If there are distinct physical reasons why robust storage and manipulation of
classical information is possible, an obvious question to ask in the quantum
information sciences is whether we can mimic these effects in the quantum
domain.  Do there exist, or can we engineer, physical systems whose physics
ensures that the robust storage and manipulation of quantum information is
possible?  In this section, we will present evidence, in the form of a mean
field argument, that a Hamiltonian related to the three-dimensional subsystem
code might be exactly this type of system.

\subsection{Self-Correcting Quantum Memories}

The traditional approach to building a robust fault-tolerant quantum computer
imagines building the computer using a complex microarchitecture of quantum
error correcting fault-tolerant procedures.  This poses a severe technological
overhead of controlling thousands of qubits in a complex manner, simply to get
a single robust qubit.  Kitaev\cite{Kitaev:97a,Kitaev:03a} was the first to
suggest that an alternative, less complex, method to constructing a
fault-tolerant quantum computer might be possible.  Kitaev showed that there
exists a quantum error correcting code, the toric code, which is the degenerate
ground state of a certain four body spatially local Hamiltonian on a
two-dimensional lattice of qubits.  Kitaev imagined encoding quantum
information into the ground state of this system and then, because there is an
energy gap between the ground state of this system and the first excited state
and because the errors which will destroy quantum information consist of error
which scale like the size of the lattice, this quantum information would be
protected from decoherence due to the environment as long as the temperature of
the environment was sufficiently low.

It is important to note that the Hamiltonian implementation of Kitaev's toric
code (by which we mean encoding information into a physical system governed by
the four-body Hamiltonian associated with the toric code), while providing a
mechanism for the robust storage of quantum information, does not provide a
full fault-tolerant method for quantum computation.  The reason for this is
that during the implementation of the physical processes which manipulate the
information encoded into the ground state of the system, real excitations will
be created which will disorder the system. This distinction has been confused
in a manner because the toric codes {\em can} be used to construct a
fault-tolerant quantum computer, but only with the aid of external quantum
control which serves to identify and correct errors which occur during the
manipulation of the information encoded into the system (see for example
\cite{Dennis:02a}).

The idea of a self-correcting quantum memory is to overcome the limitations of
Kitaev's original model by constructing a physical system whose energy levels
not only correspond to a quantum error correcting code (in our case a subsystem
code), but which also uses the energetics of this system to actively correct
real errors created when the quantum information is being manipulated.  Thus,
while in the toric codes, a single real error on the system can create
excitations which can disorder the system, in a self-correcting system, a
single real error on the system cannot disorder the system.

In order to explain the distinction of a self-correcting memory from the
original toric code we will compare the situation to that of the one
dimensional and two-dimensional classical ferromagnetic Ising model.  These
models will be analogous to the toric code Hamiltonian model and a
self-correcting Hamiltonian model, respectively.

Recall that in a ferromagnetic Ising model one takes a lattice of classical
spins and these are coupled by Ising interactions between the neighboring spins
via a Hamiltonian
\be
H=-{J \over 2} \sum_{\<i,j\>} s_i s_j,
\ee
where $s_i \in \{\pm 1\}$ are the spin variables, the sum $\<i,j\>$ is over
neighbors in the lattice, and $J>0$.  Notice that the ground state of this
Hamiltonian corresponds to the uniform states $s_i=+1, \forall i$ or
$s_i=-1,\forall i$.  These, of course, are also the two codewords for a
classical redundancy code.  Thus we can imagine that we encode classical
information into the ground state of this Hamiltonian in direct analogy to the
way in which information (but quantum this time) is encoded into ground state
of the toric code.  Errors on the Ising codes are just bit flips.  From hereon
out when we refer to the Ising model we will implicitly be discussing the
ferromagnetic Ising model.

Recall some basic properties of the one and two-dimensional Ising models (see,
for example \cite{Plischke:94a}).  We begin by discussing the thermal
equilibrium values of the total magnetization,
\be
M=\sum_i s_i,
\ee
of these models.  In one dimension, for any $T>0$, the total magnetization of
the Ising model vanishes in thermal equilibrium, whereas for the
two-dimensional Ising model, the magnetization is zero above some critical
temperature $T_c$ and below this temperature, two magnetizations of equal
magnitude and opposite sign are maintained.  Since the magnetization is a
measure of the information recorded in the redundancy code, we see that if we
encode information into the ground state of the one dimensional Ising model and
this system is allowed to reach thermal equilibrium, then this information will
be destroyed.  On the other hand, for the two-dimensional Ising model, if we
encode information into the ground state and the system is below the critical
temperature $T_c$, then this information will be maintained.  Above $T_c$, like
the one-dimensional Ising model, the information will be destroyed.  From the
point of view of storing the information in the thermal states of these models,
the two-dimensional Ising model is a robust medium, but the one-dimensional
Ising model is not.

But what about the properties of the Ising models on the way to reaching
equilibrium (i.e. during the time evolution with the environment)? In the one
dimensional case we find that the system will generically (depending on the
exact method of relaxation) take a time which is suppressed like a Boltzman
factor $e^{-J/T}$.  Thus at low enough temperature, we can encode information
into the ground state of the one dimensional Ising model and it will be
protected for a long time.  While the scaling of this decay rate is favorable
in the temperature T, this type of approach is different from what is done in
standard error correction where larger redundancy can be used to overcome
errors without changing the error rate (as long as that error rate is below the
threshold for the error correcting code.)

What happens for the time evolution of a two-dimensional Ising model?  If we
start the system in one of the redundancy code states, then far below $T_c$ the
system will relax quickly to the closest of the two equilibrium states with a
large total magnetization.  As we raise the temperature closer to $T_c$, this
relaxation will slow down.  Above $T_c$ the picture is similar to that of the
one dimensional Ising model that if we are close to $T_c$, then the relaxation
to vanishing magnetization is suppressed like $e^{-J/(T-T_c)}$.

What are the main reasons for the differences in the ability to store
information in the one and two-dimensional Ising models?  A rough heuristic of
what is happening is that in the two-dimensional Ising model, the errors
self-correct\cite{Barnes:00a}.  Consider starting the one dimensional Ising
model in the all $s_i=+1$ state.  Now flip one of the spins at the end of the
chain. This will cost an energy $J$.  Flipping the neighbor of this spin will
then cost no energy. Proceeding along the chain in this manner one sees that
one can expend energy $J$ to turn the system from the codeword all $s_i=+1$
state to the all $s_i=-1$ state.  Thus the environment need only supply this
energy to disorder the system.  However, in the two-dimensional Ising model,
something different happens.  Suppose again that we start in the all $s_i=+1$
state. Here if we flip a single spin (say on the boundary of the lattice) then
the energy required to flip this spin is $J$ times the number of bonds this
spin has with its neighbors.  Now flipping a neighbor will cost energy: in the
two-dimensional model the energy cost of flipping a connected domain of spins
is proportional to the perimeter of this domain.  Since to get from the all
$s_i=+1$ codeword to the all $s_i=-1$ codeword we need to build a domain of
size the entire lattice, we see that we will require at least an energy times
the size of the lattice to disorder the system.  Suppose, now that errors are
happening at some rate to all of the spins in the Ising models.  In the one
dimensional model, once one creates a single error, there is no energy barrier
to disordering the system.  In the two-dimensional model, however, there is now
an energy barrier. In particular, the system coupled to its environment will
not only cause errors, but will also cause errors to be corrected by shrinking
the domains of flipped errors.  As long as the error rate is not too strong
(which corresponds loosely to being below the critical temperature $T_c$) the
pathways that fix the error will dominate the actual creation of errors.

Thus we see that a two-dimensional Ising model operating below the critical
temperature is performing classical error correction on information stored in a
redundancy code.  In the one-dimensional Ising model and in the two-dimensional
Ising model above the critical temperature, there is suppression due to a
Boltzman factor, but there is no self-correction (or the self-correction is not
fast enough) and the information stored in the redundancy code is destroyed.

The two-dimensional anyon models of topological quantum computing and
variations\cite{Kitaev:97a,Ogburn:99a,Dennis:02a,Kitaev:03a,Wang:03a,Freedman:01a,Nayak:01a,Brink:03a,Freedman:03a,Freedman:03b,Hamma:04a,Freedman:04a,Freedman:04b,Freedman:05a,Kiteav:05a},
including Kitaev's toric model, all share the property with the one dimensional
Ising model that the system can disordered using only an energy proportional to
the gap in the Hamiltonian.  (The models in \cite{Hamma:04a} contain errors
similar to those in the two dimensional Ising model for certain types of
quantum errors, but not for both phase and bit flip errors.) This can provide
protection via an exponential suppression due to a Boltzman factor, but this
does not provide indefinite correction. The idea of a self-correcting quantum
memory, however is to mimic the two-dimensional Ising model.  In particular
below a critical temperature, quantum information stored in the system should
persist even when the system is in thermal equilibrium and further the system
should have a mechanism whereby real errors are corrected by the energetics of
the system faster than the real errors occur when operating below the critical
temperature.

Finally, we note that there is one system which is widely suspected to be a
self-correcting quantum memory.  This is a version of the toric code on a
four-dimensional lattice\cite{Dennis:02a}.  The problem with this model is that
it exists in four dimensions and that it requires greater than two-qubit
interactions in order to implement and is therefore not realistic for practical
implementations. Our original motivation for considering the subsystem codes
presented in this paper was to obtain a self-correcting system in the realistic
setting of three or lower dimensions and using two-qubit interactions.

\subsection{The Three-dimensional Hamiltonian}

Next we turn our attention to a system which may be a self-correcting quantum
memory.  We provide evidence for this by showing that in a mean field
approximation this Hamiltonian has properties for the expectation value of its
energy which is similar to the energetics of the two-dimensional Ising model.

Consider the following Hamiltonian on a cubic lattice of qubits constructed
exclusively from elements of ${\mathcal T}_3$,
\be
H=-\lambda \sum_{i,j=1}^n\sum_{k=1}^{n-1}\left( X_{k,i,j} X_{k+1,i,j} +
X_{i,k,j} X_{i,k+1,j} + Z_{i,k,j} Z_{i,k+1,j} + Z_{i,j,k} Z_{i,j,k+1} \right),
\label{eq:ham3d}
\ee
with $\lambda>0$.  This Hamiltonian consists of Ising couplings along the
direction $x$ in the $xy$-plane and along the direction $z$ in the $yz$-plane.
As in the two-dimensional Hamiltonian, Eq.~(\ref{eq:ham2d}), we can use the
fact that $H$ is a sum of operators from ${\mathcal T}_3$ to block diagonalize
$H$ with respect to the subsystem structure of our three-dimensional operator
quantum error correcting subsystem:
\be
H=\bigoplus_{s^X,s^Z} H_{s^X,s^Z} \otimes I_2.
\ee
Information can then be encoded into the $I_2$ subsystem.  We would like to
show that if we perform such an encoding, then the information stored in this
subsystem will be protected from the effect of general quantum errors in a
manner similar to that of the two-dimensional Ising model.

\subsection{The Mean Field Argument}

Ignore, for the moment the boundary conditions for the Hamiltonian and suppose
that the ground state of Eq.~(\ref{eq:ham3d}) has the following properties for
the expectation values of the Ising bonds in the system,
\bea
\<X_{k,i,j} X_{k+1,i,j} \>_G=c_{xx} \nonumber \\
\<X_{i,k,j} X_{i,k+1,j} \>_G=c_{xy} \nonumber \\
\<Z_{i,k,j} Z_{i,k+1,j} \>_G=c_{zy} \nonumber \\
\<Z_{i,j,k} Z_{i,j,k+1} \>_G=c_{zz}, \nonumber \\
\eea
where $c_{\alpha \beta}>0$.  In such a phase, the ground state energy is given
by
\be
E_G=\<H\>_G=-\lambda n^2 (n-1) (c_{xx}+c_{xy}+c_{zy}+c_{zz}).
\ee
Now consider the effect of a single Pauli error on a single qubit in the
lattice (assume this is away from the boundary).  For example, consider a Pauli
$X$ error.  Using the fact that $X$ commutes with Ising bonds oriented along
the $x$ direction but anticommutes with Ising bonds oriented along the $z$
direction, it then follows that expectation value of the energy of the system
changes to
\be
E_1=\<H\>_1= E_G+ 2 \lambda (c_{zy}+c_{zy}+c_{zz}+c_{zz}),
\ee
where we have separated out each term arising from each of the four $z$
direction Ising bonds which connect to the lattice site where the $X$ error has
occurred.  Thus we see that a single bit flip error on the ground state will
cause the expectation value of the energy to increase.  Now consider the effect
of applying a second Pauli $X$ error which is a nearest neighbor to the
original spin in the same $yz$ plane.  Now the Ising bond between these two
errors does not contribute to the change in the expectation value of the ground
state, but all of the $z$ direction Ising bonds around the perimeter of the two
flipped spins do contribute.  Thus, for example, if the spins are neighbors
along the $y$ direction, the expectation value of the energy of this new state
is
\be
E_2=\<H\>_2=E_G + 2 \lambda (2 c_{zy}+4c_{zz}).
\ee
Generalizing the above argument we see that a connected domain of Pauli $X$
errors in an $yz$ plane will result in an energy increase proportional to the
perimeter of the domain.  A similar argument will hold for Pauli $Z$ errors in
an $xy$ plane, but now the Ising bonds along the $x$ direction will contribute
to the change in expectation value, and those along the $z$ will not
contribute.

Suppose that a general error $P(a,b)$ occurs on the ground state of $H$. Then
in each plane $yz$ plane the part of the error coming from $a$ will produce
excitations whose expectation of the energy scales like the perimeter of
domains of errors in $a$ and similarly for each $xy$ plane, but now for the $b$
component of the errors.  From this argument we see that, at least for the
expectation value of the energy, the system looks very similar to the
two-dimensional Ising model.  But now instead of only bit flip errors, more
general quantum errors produce changes in energy of the system which are
proportional to the perimeter of the erred domain.  Thus we argue that this
provides evidence that the model we have presented will be self-correcting. For
the same reasons that the two-dimensional Ising model will not disorder the
classical information stored in a redundancy code (i.e. since our errors
require (expected) energy proportional to the perimeter of the erred domain) we
expect the quantum information stored in our system will not disorder up to
some critical temperature.

Of course, the evidence we have provided is based on numerous assumptions
arising from our mean field model. First of all we have ignored boundary
conditions. It is possible that certain edge states could disorder the system.
Secondly we have only made arguments about the expectation value of the energy
after error have occurred to the ground state. This doesn't give us concrete
information about the energy level structure of our Hamiltonian. It could be
that while the expectation value scales like the perimeter, there are actually
error pathways whose energetics are much less favorable.  Third we have assumed
the existence of a phase with the desire expectation values of the bond
energies.  This phase may not exist, i.e. it may be that in the thermodynamic
limit, the expectations values all vanish.  This would totally invalidate the
mean-field argument we have given above. Given these caveats our mean field
argument only suggests that the system will be self correcting. Clearly
rigorously establishing whether our memory is self-correcting is a challenging
open problem.

\subsection{The Quantum Error Correcting Order Parameter}

In the Ising model, an indication that the information stored in a redundancy
code is still there after thermalization is the persistence of the total
magnetization of the system. In particular, for the two-dimensional Ising model
at a temperature between zero and the critical temperature, the magnetization
in equilibrium is never exactly equal to its maximal value, $\pm n^2$.  This is
because there are always small domains of flipped spins due to the interaction
of the system with its environment.  However, a magnetization which is
different from zero may be interpreted as a measurement of the majority vote
for the redundancy code in this system.  If we are going to establish that our
quantum system is actually self-correcting, it is important to identify an
order parameter for our system which can be used to reveal the presence of
quantum information in our system. This can be done using the operator quantum
error correcting subsystem properties of the three-dimensional code.

Suppose we encode quantum information into a quantum error correcting code and
apply a number of quantum errors less than the number which the code has been
designed to correct.  We know from the theory of quantum error correction that
the encoded information in this system can be recovered by the measurement of
an appropriate error syndrome and the application of the appropriate recovery
procedure.  We can use this to construct an order parameter for any quantum
error correcting code.

A note about the nature of order parameters for quantum information before we
describe this parameter for our three-dimensional model.  In the classical
Ising model, we found that below the critical temperature there was a
bifurcation of the system into two magnetizations of equal and opposite
magnitude.   Since classical information is based upon a bit, we are not
surprised to find that such a bifurcation into two states happens.  Actually,
depending on the initial state of the system before it is thermalized, the
thermal state of the system can be any value between these two extremes.  But
if we start our system by encoding into one of the two codestates (the all $\pm
1$ state) then only the two bifurcated values will result after thermalization.
What is the analogous situation for quantum information? For quantum
information, we must show not that the bifurcation happens for a bit of encoded
information but instead for a qubit of encoded information. Since a qubit is
parameterized by the Bloch-sphere, one might expect that one needs an order
parameter with similar properties.  Such an order parameter can be constructed,
but we can get away with examining fewer parameters in order to show the
robustness of the quantum information.  In particular if we make measurements
along the $x$, $y$, and $z$ directions of the quantum information, then because
of the linearity of the density operator we can use this to show that the
information has been preserved.  In particular, we can imagine encoding into
one of the eigenstates of Pauli operators along these directions and looking at
this system after it has thermalized.  Notice now that instead of a single
order parameter, we will have three order parameters. In order to demonstrate
the self-correcting nature of our three-dimensional Hamiltonian, we will need
to show that the expectation value of these three order parameters each
bifurcate below some critical temperature.

Consider, now, the active recovery procedure for our three-dimensional
subsystem code.  We begin by measuring the $S_k^X$ and $S_i^Z$ operators. Given
these measurements, we can, as in the active recovery procedure, deduce an
appropriate recovery operator to restore the information originally encoded
into a subsystem, modulo the subsystem structure of the system.  If we were to
apply this syndrome and measure the encoded Pauli logical operators for the
code (which were given in Sec.~\ref{sec:log3d}) then this would serve as an
order parameter for our system.  We note, however that if we are simply
interested in measuring the Pauli logical operators and not in fully restoring
the information into the original subsystem encoding, we do not need to
actually apply the syndrome.  This is because the syndrome we diagnose will
either commute or anti-commute with the encoded Pauli operator we wish to
measure. Thus given the syndrome we can, instead of applying the appropriate
recovery operation, simply flip or not flip the answer we get from measuring
the encoded Pauli operator depending on the exact syndrome measured.

Now we can explicitly describe our order parameters.  Suppose we measure the
$S_k^X$ and $S_i^Z$ operators and obtain the values $s_k^X$ and $s_i^Z$ and we
measure one of the encoded Pauli operators.  Suppose, for example, that this
encoded Pauli operator is an encoded $X$ operator which we measure by measuring
all of the Pauli $X$ operators in a fixed $xy$ plane. Given the $s_k^X$ we can
deduce whether an even or an odd number of Pauli $Z$ errors will need to apply
to the fixed $xy$ plane to restore the quantum information in the subsystem (if
possible).  If this number is odd, then we simply flip the value of we measure
for the encoded Pauli $X$ operator and if this number is even, we do not flip
the value for the encoded $X$ operator.  Thus we see that we measure the $x$
directional order parameter from our system by measuring the $S_k^X$ and the
encoded $X$ operator and, as a function of these values, produce a single
number representing the $x$ directional order parameter.  Similar comments hold
for the order parameters along the other cardinal directions.

\section{Conclusion}\label{sec:conc}

In this paper we have constructed a new class of quantum error correcting
procedures based upon the notion of encoding quantum information into a
subsystem.  By encoding into a subsystem, we were able to demonstrate a
recovery routine which explicitly used the subsystem structure.  The
three-dimensional code we constructed was shown to be related to a
three-dimensional spin lattice system which we gave evidence for being a
self-correcting quantum memory.  We close by remarking on some open problems
for this three-dimensional system and some thoughts about future directions for
constructing self-correcting quantum systems.

The first open question concerns the implementation of our model in a physical
system.  A particularly promising system for such an implementation is with
ultracold atoms trapped in an optical lattice\cite{Deutsch:00a}.    Duan,
Demler, and Lukin\cite{Duan:03a} showed how to simulate a large class of
spin-spin interactions for these systems.  An open question is whether their
techniques allow one to implement our three-dimensional anisotropic spin-spin
Hamiltonian. Of particular concern is the magnitude of the spin-spin coupling
which one can achieve in these models. This will directly effect the critical
temperature for any self-correction that occurs in the system.  A further
concern for the physical implementation in an optical lattice is the ability to
measure the syndrome operators $S_k^X$ and $S_i^Z$ along with appropriate
logical Pauli operators. Finally one would like to understand how to produce an
effective controlled-not coupling between two such encoded lattices. Solutions
to all of these problems would allow one to propose an experiment in which a
self-correcting quantum memory could be demonstrated.

A second open question is, of course, whether our three-dimensional system is
indeed self-correcting.  Noting that the Hamiltonian for this system does not
possess a sign-problem, one approach to verifying this question would proceed
by performing quantum Monte Carlo simulations of this system.  The order
parameters we have described could then be simulated at finite temperature and
evidence for self-correction could then be examined.  Another promising
approach is to use recent new ideas in the density matrix renormalization
group\cite{Vidal:03a,Verstraete:04b} to simulate the thermal properties of this
system.

Another important question is whether one can design a self-correcting quantum
systems in two dimensions.  This would be particularly desirable if one wishes
to physically implement the self-correction in a solid-state system.

Finally a large open question is what role operator quantum error correcting
subsystem codes can play in quantum information science.   What do other such
subsystem codes look like?  Since these codes have important properties due to
the fact that they exploit degenerate quantum codes, can subsystem codes be
used to beat the quantum Hamming bound\cite{Gottesman:97a}?  Further, results
which relied on showing that there were no subspace codes with certain
properties (for example, as in \cite{Reimpell:05a}) need to be reexamined in
light of the existence of operator quantum error correcting subsystem codes.

\begin{acknowledgements}

This work was supported in part by the DARPA QuIST program under grant
AFRL-F30602-01-2-0521.  D.B. acknowledges the support of a postdoctoral
fellowship at the Santa Fe Institute, during which some of initial work on this
paper was carried out.  D.B. also thanks Kaveh Khodjasteh, Daniel Lidar,
Michael Nielsen, and David Poulin for useful discussions.

\end{acknowledgements}

\end{document}